Probabilistic Grading and Classification System for End-of-Life Building Components Toward Circular Economy Loop

Yiping Meng[1], Sergio Cavalaro[2], Mohamed Osmani[2]

Abstract: The longevity and viability of construction components in a circular economy demand a robust, data-informed framework for reuse decision-making. This paper introduces a Multi-Level Grading and Classification System (MGCS) that combines Bayesian probabilistic modeling with scenario-based performance thresholds to assess the reusability of end-of-life (EoL) modular components. By grading components across a five-tier scale (A–E), the system supports strategic decisions for reuse, up-use, or down-use, ensuring alignment with engineering standards and sustainability objectives. The model's development is grounded in empirical data from precast concrete wall panels, and its explainability is enhanced through decision tree logic and Sankey visualizations that trace the influence of contextual scenarios on classification outcomes. MGCS addresses the environmental, economic, and operational challenges of EoL management—reducing material waste, optimizing value recovery, and improving workflow efficiency. Through dynamic feature weighting and transparent reasoning, the system offers a practical yet rigorous pathway to embed circular thinking into construction industry practices.

1. Introduction

1.1 The challenges faced during the disposal or reuse of End-of-life (EoL) building components.

When the life span of buildings has ended, most of the building components will become construction and demolition waste (C&D) with a lower material recycling rate. Reuse is different from recycling, which generates new materials after certain processes. Reuse preserves the functions of the used components after interventions like repair (Iacovidou and Purnell 2016). With the extensive use of building materials like concrete and the high environmental impact, the reuse of building components can ideally reduce the carbon footprint (Cai and Waldmann 2019a) and achieve carbon neutrality through reducing C&D waste and the generation of substitutes for primary materials and products (Cai and Waldmann 2019b). Reusing construction elements can efficiently reduce the environmental impact of building construction, especially when implemented over multiple life cycles (Fivet 2019) . Since building components with similar ages may not be in the same state (Vanier 2001), there would be possibilities that the EoL building components could be used in the same or other usage scenarios. Concrete, and exceptionally high-performance concrete, has durability qualities, which allows multiple usage cycles. Circular economy principles encourage reusing EoL products, while barriers exist in the construction sector. The biggest challenge is from the technical sector.

(1) Quality assurance: Ensuring and maintaining the quality of reused materials and components presents a formidable challenge. Scrutinizing materials for wear, degradation, or structural defects necessitates comprehensive inspection and testing protocols. The disparity in quality, owing to varied usage histories and potential latent

[1] School of Computing, Engineering and Digital Technologies, Teesside University, y.meng@tees.ac.uk

[2] School of Architecture, Building and Civil Engineering, Loughborough University, s.cavalaro@lboro.ac.uk, m.osmani@lboro.ac.uk

defects, demands robust and reliable testing methodologies, possibly leading to augmented operational overheads.(Ajayi et al. 2015; Rakhshan et al. 2020)

(2) Material Complexity and Diversification: Construction materials exhibit a vast array of properties and characteristics, shaped by varied manufacturing processes, usage histories, and exposure to different environmental conditions. Navigating through this complexity, ensuring materials are suitably allocated and utilized, post-reuse, demands technical expertise, and advanced material science applications.

(3) Deconstruction Difficulties: Unlike demolition, deconstruction with a focus on material reuse necessitates careful dismantling of structures to prevent damage to components intended for reuse. This procedure is not only labour-intensive but also requires skilled personnel adept at preserving the integrity of the materials during the deconstruction process.

(4) Standardization and Certification: Developing standardized metrics and certification protocols to grade and validate the quality and reliability of reused materials and components is an arduous task. Achieving consensus among stakeholders, ensuring the devised metrics are universally applicable, and ensuring adherence is complex and resource-intensive.

(5) Logistical Intricacies: Effective material management, from deconstruction, sorting, storage, and transportation to the final re-employment, incurs logistic complexities. Optimizing this chain to ensure timely, cost-effective, and quality-consistent availability of reused materials is a pivotal challenge.

(6) Compatibility and Interoperability: Ensuring reused materials and components are compatible with new construction methodologies and materials, safeguarding structural integrity, and adherence to modern design principles and regulatory frameworks, is intricate. This often necessitates additional modifications or adaptive strategies to align reused components with current building practices.

(7) Regulatory and Compliance Hurdles: Navigating through the regulatory landscape, ensuring reused materials and constructions adhering to them are compliant with existing building codes, safety standards, and environmental regulations, demands thorough due diligence and often complex bureaucratic navigation.

(8) Technological Adaptation: Adapting to or developing technologies that facilitate effective material testing, deconstruction methodologies, and construction practices that are conducive to material reuse requires investment, research, and skilled personnel.

Other challenges include economic impediments to costs and environmental concerns for hazardous materials faced by EoL building components. Also, the EoL components might have deteriorated to a point where reuse is not technically feasible A condition assessment can serve as a benchmark for the status classification for building components to support decisions for the interventions for the circular economy.

1.2 The Current state-of-the-art of reuse, repurpose, and recycling in Modern Methods of Construction (MMC)

Modern Methods of Construction (MMC) broadly encompass techniques that involve off-site construction and later assembly on-site. MMC are being used in the construction industry, particularly for housing, as they potentially represent savings in time and materials, and provide higher standards of quality than more conventional methods of construction. While

MMC offers several advantages, such as speed, quality control, and reduced waste, the challenges and advancements related to their reuse, repurposing, and recycling are essential from a sustainability perspective. However, it is vital to consider the challenges and advancements related to reusing, repurposing, and recycling MMC from a sustainability perspective.

(1) Reuse:

Deconstruction Over Demolition: Increasingly, the industry is focusing on deconstructing MMC components rather than outright demolition. This approach preserves components for direct reuse in other projects.

Design for Disassembly (DfD): Many MMC components, especially volumetric units, are designed with disassembly in mind (Akinade et al. 2017). This feature ensures these units can be reused multiple times across different projects.

Digital Twins: These are digital representations of physical assets. They can track the condition and performance of MMC components over time (Boje et al. 2020). Using digital twins, decisions related to component reuse can be more data-driven (Sepasgozar et al. 2020).

(2) Repurpose:

Modular Flexibility: Some MMC components, like volumetric units, are designed to be flexible (Hořínková 2021). For instance, a residential unit module could be repurposed as a commercial space or vice-versa (Anon n.d.-c). If the design of different building types has a similar modulus, for example, the panel of an apartment can be applied to several houses (Huuhka et al. 2015).

Component Refurbishment: Before repurposing, MMC components might undergo refurbishment to meet the standards of their new purpose. This might involve updating insulation, altering internal layouts, or replacing finishes (Cumo et al. 2022).

1.3 The need for a grading and classification system.

The implementation of a well-defined grading and classification system can play a vital role in upholding the quality, safety and structural integrity of EoL building components especially for load-bearing ones. The grading and classification can effectively tackle the technical challenges with quality and safety assurance of reusability and contribute towards fostering a more sustainable and circular economy in the construction industry (Foster, Kreinin, and Stagl 2020). The grading tool helps construction stakeholders assess repair and maintenance needs for EoL building components (Faqih and Zayed 2021). A comprehensive grading and classification system can help address several challenges, such as:

(1) Condition assessment and service-life calculations: It is important to assess the condition of building components at the end of their life cycle and estimate their remaining service life to determine their potential for different circular interventions like reuse, repurpose or recycle (Suchorzewski, Santandrea, and Malaga 2023). A grading system can standardise the assessment process for multi-cycles of circular economy and clarify the quality and durability of components after EoL.

(2) Grading and classification for reuse and recycling: A grading and classification system can help identify the most suitable reuse or recycling options for EoL building components, considering factors such as material composition, environmental impact, and potential applications (Figl et al. 2019). This can facilitate better decision-making and promote the efficient use of resources.

(3) Deconstruction recommendations and strategies for circular interventions: A grading and classification system can inform deconstruction strategies, ensuring that valuable components are recovered, reused, or recycled effectively (Bertino et al. 2021). This can help minimise waste and reduce the environmental impact of building demolition.

(4) Environmental impacts of EoL building components: Understanding the environmental impacts of EoL building components is essential for making informed decisions about their disposal or reuse (Khasreen, Banfill, and Menzies 2009). A grading and classification system can help quantify these impacts and guide the selection of the most environmentally friendly options.

By implementing a grading and classification system for EoL building components, the construction industry can better manage the end-of-life phase of buildings, reduce waste, and contribute to a more sustainable and circular economy.

1.3 Research objectives

However, the shift from liner construction mode to circular one focuses more on the recylce at the material level, little attention is paid to how to develop the circular mode for MMC products and the manufacturing process to increase the reusability. To fill this gap, this research aims to develop and validate a comprehensive Multi-Level Grading and Classification System (MGCS) for (MMC) components, facilitating their reuse, repurposing, and recycling in a circular economy context. The objectives of the research are as follows:

(1) To identify and characterise the essential features influencing the performance and longevity of MMC components, focusing on their suitability for reuse, repurposing, and recycling.

(2) To formulate a grading system based on the identified features, allowing for an assessment of MMC components from 'very good' to 'bad' (grades A to E) based on national standards and guidelines and adapt it for MMC specifics.

(3) To develop a classification system that determines the most appropriate circular interventions for MMC components—whether reuse, up-use, down-use, or material recycling—based on their graded condition and potential application scenarios.

(4) To ensure practicality and effectiveness in real-world applications, validate the proposed MGCS through a series of case studies focusing on key MMC components such as wall panels and volumetric units.

2. Development of Grading System

According to ISO 20245 2017, which is the first global technical specifications on second-hand goods for cross-border trade, the evaluation of the used goods is based on several acceptance criteria, namely safety, quality, production information and usage requirements (Anon 2017). According to the criteria requirement, there are four classifications for the condition of the used goods in "A", "B", "C", "D" rankings.

"A": "Very good" condition. Class A products should have all their primary and secondary features available (operational). In addition, operating instructions, maintenance manuals, care instructions and parts manuals should be provided, preferably in the language of the consignee.

"B": "Good" condition. Class "B" products should have all their primary and most secondary features available (operational). Where practical, operating instructions, maintenance manuals, care instructions and parts manuals should be provided, preferably in the language of the consignee.

"C": "Acceptable" condition. Class "C" products should have most of their primary and secondary features available (operational).

"D": "Unfit" condition. Class "D" products have most primary and secondary features unavailable (non-operational) and should be traded only to extract parts for aftermarket needs.

There is no commonly used rating or grading system for building construction for EoL components. Some building rating systems are applied to different building types, like hospital or residential buildings, for monitoring defects (Straub 2009) or maintenance of the existing buildings (Abbott et al. 2007; Salim and Zahari 2011). The lowest grade of the EoL building component means that the components can not be applied at the product level and can only be recycled into material. Based on the four ranking levels of ISO 20245 2017 and commonly used rating scales for buildings (Faqih and Zayed 2021), one additional grading level – "bad" is added, which means all the features are unavailable and the component can only be recycled.

Figure 1 displays the multi-level grading and classification (MGCS) system that encompasses the entire cradle-to-cradle life cycle. The MGCS is situated between the C1-C4 End of Life stage and the D reuse stage. It connects the grades of the EoL component conditions to various circular interventions classifications. The EoL building component undergoes a five-level grading system before being categorised under different circular interventions that dictate various "reuse" techniques like repurposing. There are five different classes that require varying levels of intervention in order to ensure reusability. The meanings of the five classes are as follows:

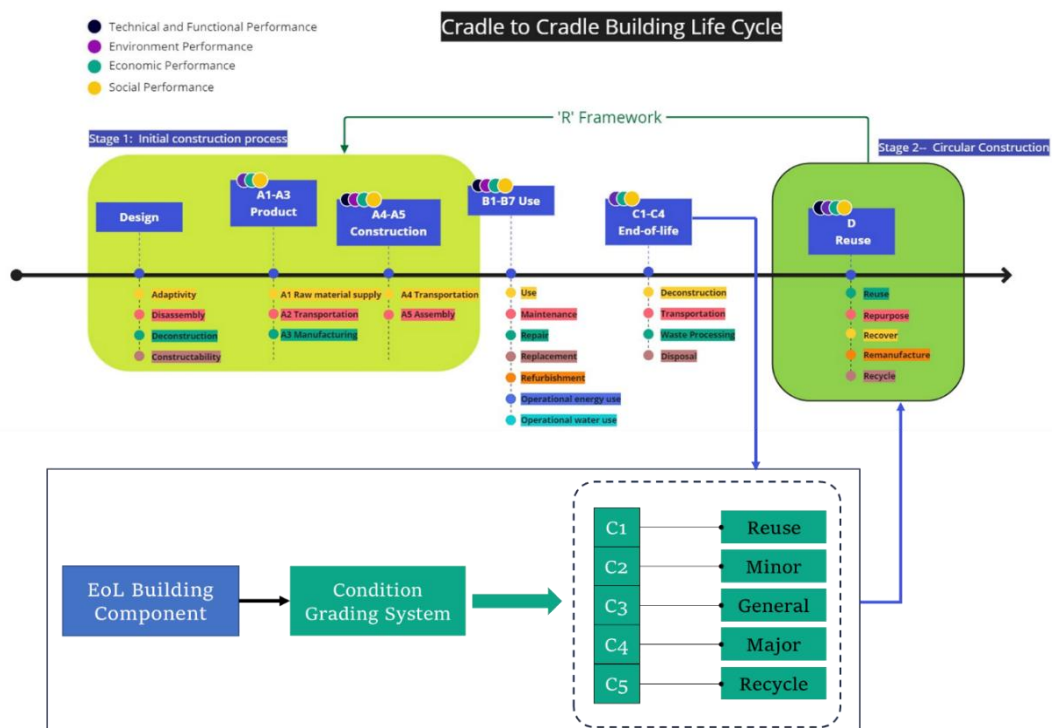

Figure 1 Grading and Classification System in Cradle-to-Cradle Life Cycle

Class 1 (Reusable): Components in very good condition can be reused without any or minimal refurbishment interventions.

Class 2 (Minor Repair and Refurbishment): Components that are in good condition and can be reused with minimal refurbishment

Class 3 (General Maintenance): Components that require refurbishment before reuse.

Class 4 (Major Repair and Refurbishment):

Class 5 (Recyclable): Components that can't be reused but can be recycled to extract material or elements.

According to the performance requirements in ISO 20245 2017 and the service life of the building in 15686-7:2017 (BSI British Standards n.d.), the performance level and the change through the whole life cycle are demonstrated in Figure 2. The performance degree thresholds are added to assess EoL product performances, ensuring they meet requirements for multi-cycle building components.

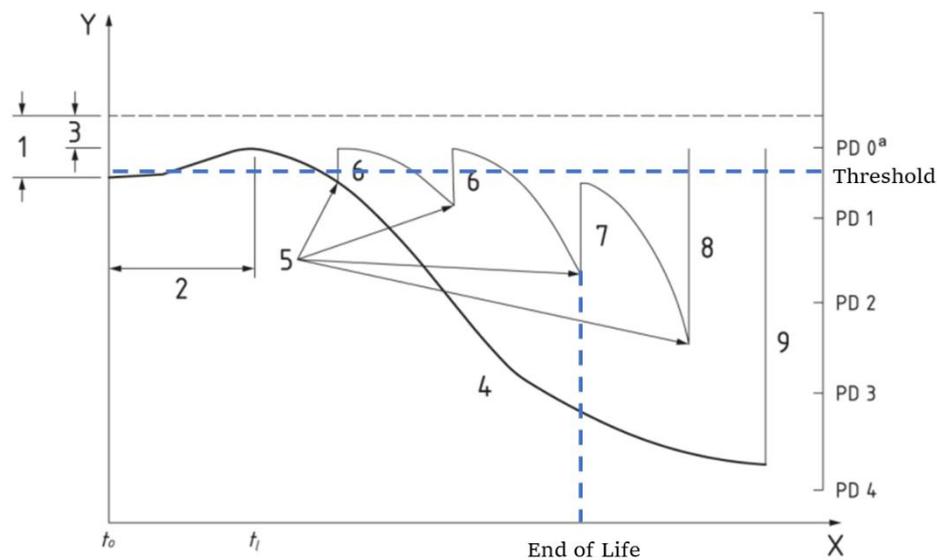

Y: Quality/Function  
X: Time  
PD: Performance Degree  
$t_0$: Time of initial "as built"  
$t_i$: Time at the start of "in use" stage  
1: Expectation  
2: Commissioning  
3: Initial Performance Gap  
4: Performance without preventative actions  
5: Limit stats  
6: Preventative and Periodic Maintenance  
7: Refurbishment or Repair  
8: Replacement  
9: Repair  

Figure 2 Revised Whole Life cycle performance of construction based on ISO 20245 2017 Part 7

The further developed workflow from the one shown in Figure 1 to operate the rating for EoL product performances to support the classification of different levels of circular interventions is shown in Figure 3 and Figure 4 to ensure the performance degree and the threshold meet the expectation of the building components. The EoL component would go to the grading system composed of inspection, usage determination and regulation three parts. The inspection is conducted to assess the building's performance from quality, health, safety and stability perspectives by referencing the building components' standards and regulations. Non-Destructive Testing (NDT) instruments will support the visual inspection, which is one of the most widely used assessment methods (Faqih and Zayed 2021), to provide more reliable evaluations. After conducting an inspection, we define the usage scenarios (e.g. internal wall for concrete panel) and essential features (fire resistance) in accordance with industry standards. The grading system rates the EoL products for each feature $i$ to get the collection of the grades of several features $G = \{GF_1, GF_2, \ldots GF_i, \ldots GF_n\}$. Based on the collections for the grades $G$, we can get an overall grade for different usage scenarios with certain probabilities $P_j - U_i = \{P - U_1, P - U_2, \ldots P - U_i, \ldots P - U_N\}$, $(i = 1, 2, \ldots, N, j = A, B, C, D, E)$. $P_j - U_i$ denotes the probability of grading as $j$ under the $i$th usage. And $P - U_1 = (P_A - U_1, P_B - U_1, P_C - U_1, P_D - U_1, P_E - U_1)$ is the collection of probability of grading from A to E under the first usage. Other probability collections are the same as $P - U_1$.

$$P - U_2 = (P_A - U_2, P_B - U_2, P_C - U_2, P_D - U_2, P_E - U_2)$$

$$P - U_i = (P_A - U_i, P_B - U_i, P_C - U_i, P_D - U_i, P_E - U_i)$$

$$P - U_N = (P_A - U_N, P_B - U_N, P_C - U_N, P_D - U_N, P_E - U_N)$$

Figure 5 demonstrates an example of these probabilistic results. The grade collection $G$ including grades for the selected features $\{GF_1, GF_2, \ldots, GF_i, \ldots, GF_N\}$. Based on the defined usage scenarios, which demand various performance features, for example usage 1 needs features $F_1, F_2, F_3$, and $F_4$ to meet the performance threshold. We can get the grades for the usage 1 as $\{GF_1, GF_2, GF_3, GF_4\}$ to get the overall grade from A to E with probabilistic. For example, there is a 20% probability that the grade for usage 1 will be A, which suggests that the most likely grade for this EoL product is B. Similar processes for other usage scenarios. With the usage-determined feature-based grading process, we can provide the evaluation for the performances of EoL products for different usage scenarios, which makes the results more robust.

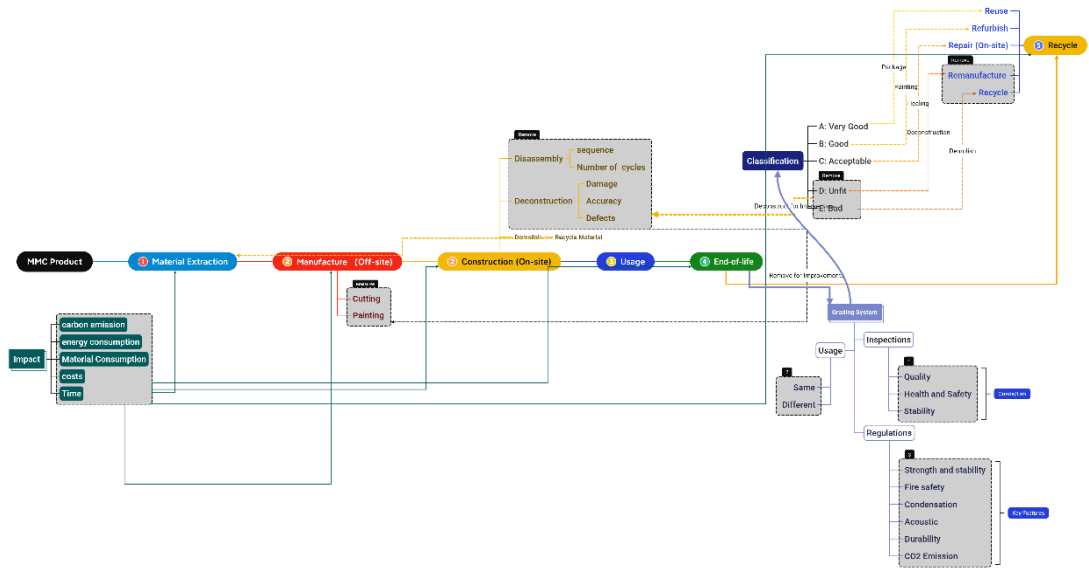

Figure 3 Workflow of Grading and Classification for EoL Product

After the grading performances for different usage scenarios, the next step is determining whether this EoL product would be reused as the same usage, as shown in Figure 4. Whether the usage is the same will affect the circular intervention levels. For example, a panel is used as a façade in the initial usage scenarios. After the life span, the inspections and the grading processes are conducted with overall grades for different usage scenarios, including the reuse as the façade, the up-cycle as the load-bearing wall and the down-cycle as the cladding. We can have the grades for the three usage scenarios according to the probabilistic grading results, supporting the following decisions: reuse as façade, upcycle as load-bearing walls, or downcycle as claddings. After the decision is made, it comes to classifying the level of circular interventions, depending on the decisions for the usage scenarios.

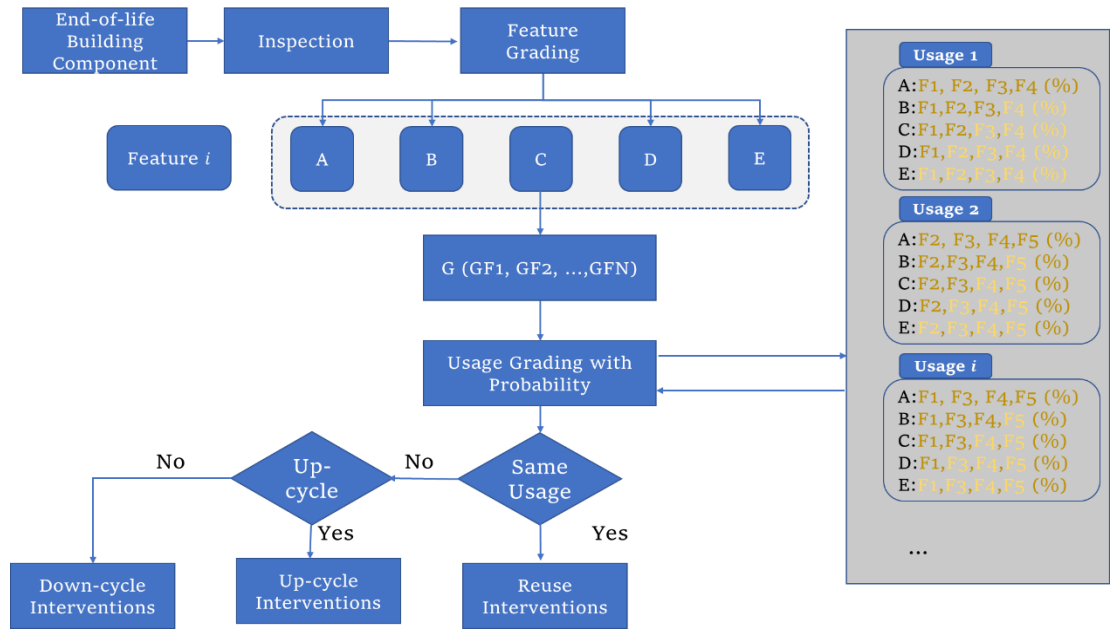

Figure 4 Grading system for EoL product based on features

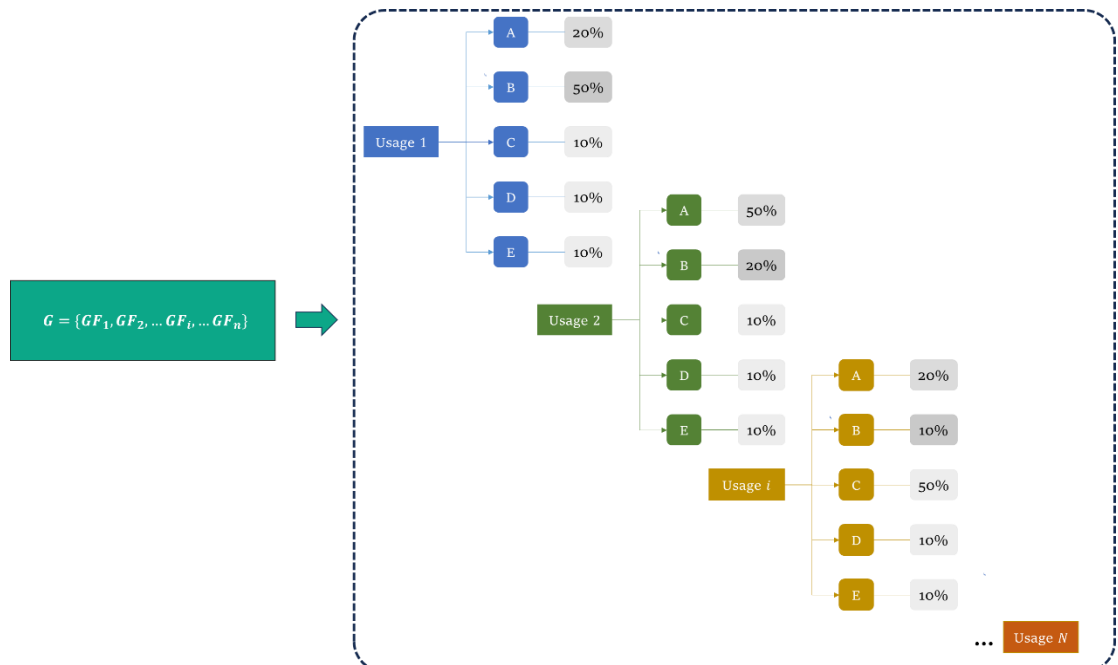

Figure 5 Probabilistic Grades under Different Usage Scenarios

For the classification for the circular intervention levels for different usage determinations, the workflows are shown in Figure 6-Figure 8. Figure 6 shows that based on the different grades for the EoL components, there would be different circular interventions in different level to ensure the components can meet the performance threshold to be reused. The ultimate outcome of the decision regarding reuse would either be to reuse the item or to recycle it. In more details, if the component is graded as A, then after processes like label and package, the component can be reused in the same usage. If a component is graded as B, it can be refurbished with minor circular interventions to improve its quality to grade A, making it suitable for reuse. The similar situation for grade C and grade D. If the component is graded as E, which means most of the features of the component is not available, then it will undergo the demolition to recycle at the material level.

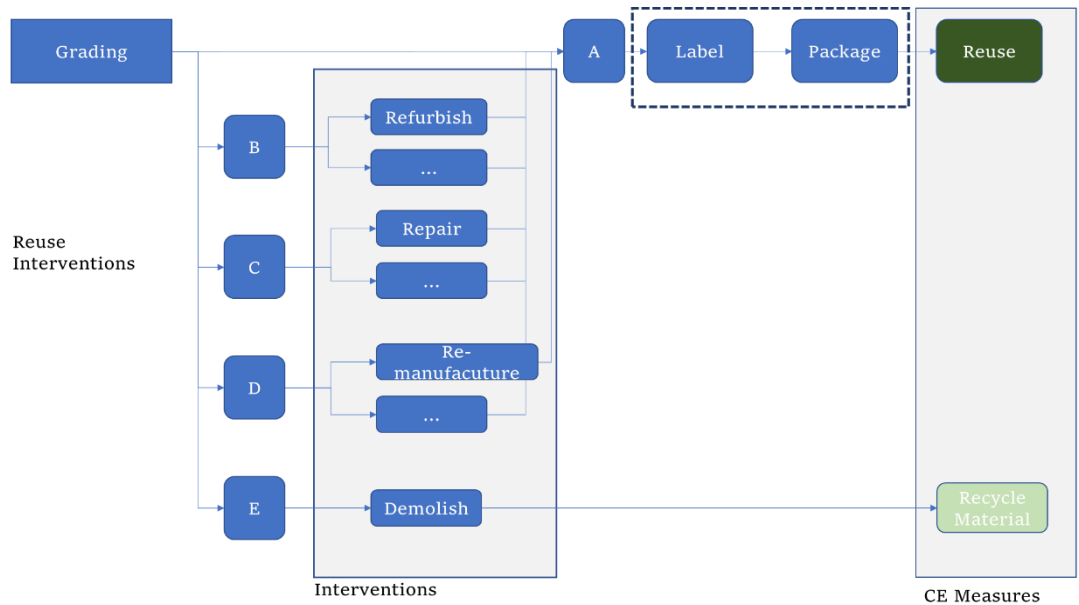

Figure 6 Circular Interventions for Reuse

The process of upcycling involves creating new uses and recycling materials, similar to circular reuse interventions. The hypothesis is that as demand for performance decreases, the number of usage scenarios increases. Supposing we have an EoL component and initially used in $i$th usage scenario, with different grade results, we can have different level of circular interventions to improve the product to upgrade the usage scenarios.

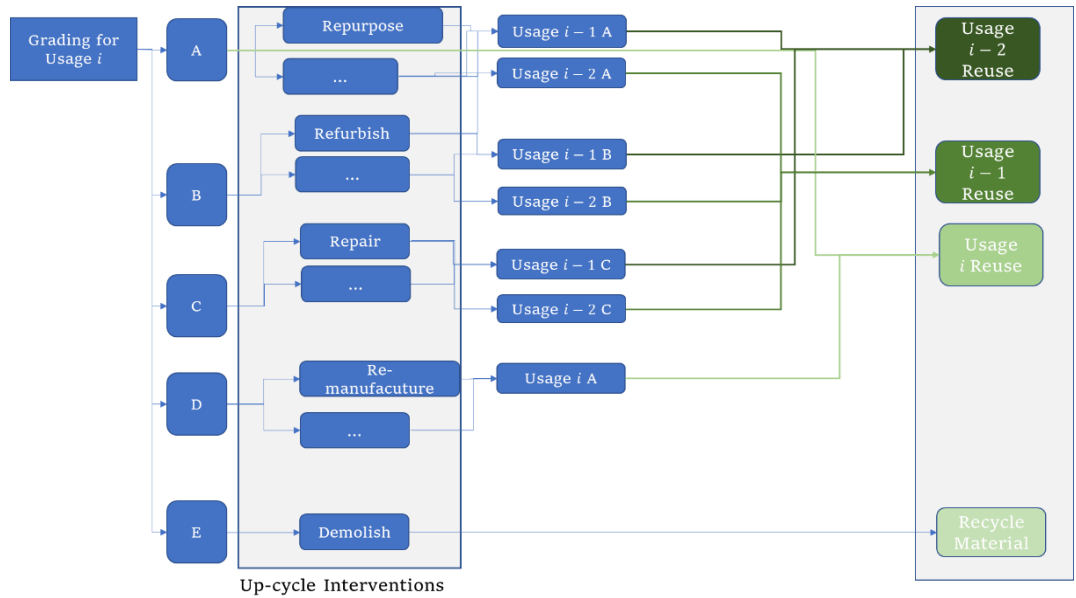

Figure 7 Circular Interventions for Upcycle

If the grade is A for usage $i$, after some process like cutting for repurpose, it can also meet the performances threshold for usage $i-1$ or usage $i-2$ as grade A to be used in usage 1 or usage 2 scenarios, which means reuse in upper level. If the EoL component is graded as B or C for usage $i$, after minor or general circular interventions, it can be improved to grade B or C for usage $i-1$ or usage $i-2$. For grade D in the $i$th scenario, it can still be maintained to grade A in the same usage scenario. However, if the grade for usage $i$ is E, it means most of the essential features are unavailable and it can only be recycled into materials.

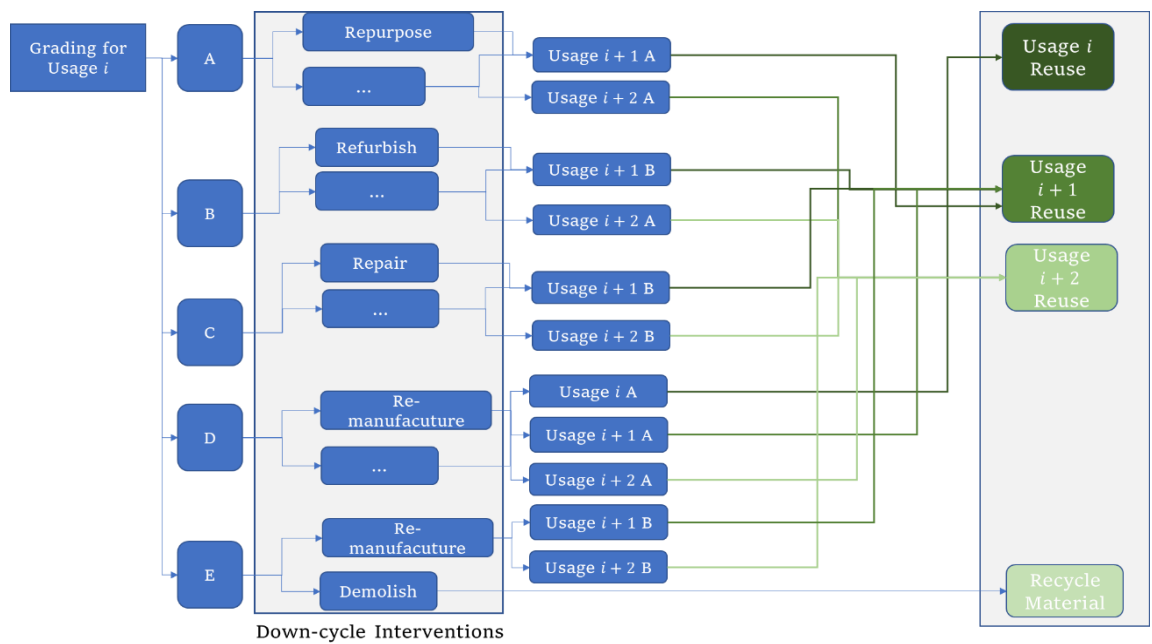

Figure 8 Circular Interventions for Downcycle

The workflow for the downcycle interventions is similar to upcycle one. The number of the usage increases as the requirement for the performance decreases. For the component that used in usage $i$ as the initial application, if it is graded as A for the $i$th usage scenario, after minor repurpose interventions like re-paint it can meet the performance threshold for grade A for usage $i+1$ and usage $i+2$, which are lower level of usage scenarios. The situation for grades B, C and D are similar, which means that after minor, general and major levels of interventions, the performance degree can be improved to grade A or B for lower level of usage scenarios. For grade E, there could be possibilities to achieve the grade B or C in usage $i+1$ and usage $i+2$ after certain major level of interventions, as the usage $i$ as the initial usage is in a higher level with better capacity and performance degree. Another situation for grade E is demolish to recycle the materials.

3. Classification of Circular Intervention Categories

Another essential part of MGCS for the EoL MMC components is the circular interventions, which aims to improve the performance degree of the components after EoL to a certain level to achieve reusability. These interventions are based on the component's types or functions and dictate what measures to take. According to the MMC definition framework, shown in 错误!未找到引用源。, seven categories of MMC processes and products happen off-site, near-site and on-site. MMC can produce products for different types of buildings like houses, low-rise (<5 story), mid-rise (6-9 story), and high-rise (10 stories and above) using different building materials (Anon n.d.-b). The MMC products includes structural and non-structural assemblies in panelised and volumetric forms. According to Construction Playbook of UK (GOV UK 2022), the Department for Business, Energy and Industrial Strategy (BEIS) and Infrastructure and Projects Authority (IPA) are developing metrics that can be commonly used to assess the performances of MMC components and buildings. There are different standards and requirements for component performance based on the various forms of MMC product categories, resulting in different types of interventions to achieve the reuse of EoL MMC components. For the performance degree assessment of MMC components, the standards specifically test or accredit for MMC products are limited. There are some accreditation schemes for MMC[3], which are not universal schemes. The existing quality assurance standards and legislations are applicable to MMC (Anon n.d.-a), like ISO, British Standards, and product certification schemes. The circular interventions aim to enhance the reusability of end-of-life (EoL) mobile manufacturing cell (MMC) products. For the inspection of the performance degree, the warranty requirements for new MMC products are selected as a reference. Generally, the technical assessment includes the detailed performance information (Anon n.d.-d):

- Structural integrity
- Performance in fire situations
- Resistance to waster penetration
- Safety in use
- Acoustic characteristics
- Thermal and movement characteristics
- Compatibility of materials (interaction between components, structural or otherwise)
- Durability and longevity of materials
- Maintenance Issues

---

[3] The most widely used accreditation for MMC are - Build Offsite Property Assurance Scheme (BOPAS) and NHBC in UK.

According to NHBC technical document for prefabricated building unit (NHBC n.d.), the performance requirements for the prefabricated components include the strength and stability, energy efficiency, fire resistance, durability and safety in use. Therefore, the inspection for the EoL MMC components should follows these performance requirements according to the specific usage scenarios.

When it comes to preserving and improving the performance of EoL MMC components through circular interventions, the main variation lies in the level of intervention. This indicates that components with different grades may have the same measures, such as refurbishment, but the material consumption or duration may differ. The classification of the level of interventions depends on these factors and we use the sustainability impact as the benchmark to set the boundaries of different levels: minor, general, and major.

### 3.1. Bayesian Classifier for the Classification

As the EoL product is graded based on different features, to obtain the overall grade for the product, the Bayesian classifier is applied to use probability method to get the results, shown in Figure 5.

Bayesian Classification is a statistical method for predicting an object's class based on one or more features. In the context of grading and classifying EoL MMC components, Bayesian Classification can be employed to systematically and predictively classify units into various grades (A-E) based on multiple features or characteristics.

(a) Defining Classes:

Class Definitions: Define the classes clearly, for example, Grade A to Grade E, based on the predetermined criteria and descriptions for various features like structural integrity, material health, safety compliance, aesthetics, and more.

(b) Defining Features:

Feature Definitions: Define the features that impact the classification, such as structural integrity, aesthetics, material health, etc.

Feature Values: Describe or quantify the status/condition of these features. It might be binary (damaged/not damaged), categorical (minor/major damage), or continuous (percentage of original load-bearing capacity).

(c) Data Collection:

Component Information: Gather information/data about EoL components, considering all defined features.

Historical Data: If available, historical data of previous assessments and interventions should also be collected and utilized.

(d) Training the Classifier:

Train with Existing Data: Use the collected data to train the Bayesian classifier, providing it with examples of EoL components and their assigned grades based on assessments.

Feature Probability: Calculate and understand the probability of particular feature conditions given a particular grade/class.

(e) Probabilistic Classification:

Probability Calculation: Use Bayes' theorem to calculate the probability of a component belonging to a particular grade, given its feature conditions.

Classification Decision: Classify the components into one of the grades based on the highest posterior probability calculated.

(f) Testing and Validation:

Test Data: Use a separate set of data (not used in training) to test the classifier's accuracy and reliability.

Validation: Evaluate the classification results against actual assessments to validate the efficacy of the classifier.

Based on the predicted grade/class, the decision for the appropriate circular interventions (refurbishment, remanufacturing, etc.) can be made. Depending on the accuracy and predictive power of the model, the adaptive strategies for EoL management can be developed. By using Bayesian Classification, the grading process can be streamlined. It allows for predictive, data-driven decision-making, which may uncover insightful patterns and relationships between different features and grading outcomes. This can lead to improved intervention strategies and the development of more resource-efficient and sustainable end-of-life practices for MMC products.

The application of Bayesian Classifier for the EoL overall grade follows the steps as:

(a) Define Prior Probabilities:

Assign prior probabilities $P(G)$ for each overall grade $G$ $(A, B, C, D, E)$ based on historical data or expert judgment. For instance:

$$P(G = A) = 0.2, P(G = B) = 0.3, P(G = C) = 0.3, P(G = D) = 0.15, P(G = E) = 0.05$$

(b) Define Likelihood:

Determine the likelihood $P(F1, F2, \ldots, F7 \mid G)$ that a component with grade $G$ would have the feature grades $F1, F2, \ldots, F7$, which might be obtained from historical data or expert estimates.

(c) Calculate Posterior Probabilities:

Using Bayes' Theorem, calculate the posterior probabilities for each grade given the observed feature grades.

$$P(G \mid F1, F2, \ldots, F7) = \frac{P(F1, F2, \ldots, F7 \mid G) \times P(G)}{P(F1, F2, \ldots, F7)}$$

Where:

$P(G)$ is the prior probability of grade G;

$P(F1, F2, \ldots, F7 \mid G)$ is the likelihood of observing the feature grades given grade G;

$P(F1, F2, \ldots, F7)$ can be calculated as: $P(F1, F2, \ldots, F7) = \sum_G (P(F1, F2, \ldots, F7 \mid G) \times P(G))$;

(d) Choose the Final Grade

Select the grade with the highest posterior probability as the overall grade.

$$G_{final} = argmax_G P(G \mid F1, F2, \ldots, F7)$$

3.2. Applying Bayesian Classifier for Multiple Usage Scenarios

Out research is based on a context where the usage scenario influences the grading, the Bayesian classifier is revised to accommodate different usages. Different usage scenarios require different number of features. For example, seven features for $U_1$, five for $U_2$ and three for $U_3$. Therefore, the probability model needs to combine the probability results with the usage scanrio. The basic principle will remain the same, but the likelihoods and potentially the priors will be influenced by the usage scenario, denoted as U (e.g., $U_1$ for residential, $U_2$ for commercial, etc.). The revised Bayesian classifier would consider not only the influence of feature grades $F$ on the overall grade $G$ but also how the usage scenario $U$ affects it. This introduces a conditional dependence of the features $F$ on the usage $U$.

Given $G$ is the grade and $F$ is the feature vector, and $U$ represents a particular usage scenario, the adjusted Bayesian classifier is:

$$P(G|F,U) = \frac{P(F|G,U) \times P(G|U)}{P(F|U)}$$

Where:

$F_i = (F_{i.1}, F_{i.2}, \ldots F_{i,N})$ is the subset for usage scenario $U_j$ $(j = 1,2,\ldots,M)$.

The approach for this adjusted Bayesian classifier for multi-usage scenario follows the steps:

(a) Define Prior Probabilities

Define the prior probability for each grade $G$ in the context of usage $U$, such as $P(G|U_j)$, considering the relevance and importance of various features for $U_j$.

(b) Define Likelihood

Define the likelihood of observing feature set $F$ given a grade $G$ and usage $U$, $P(F_i|G, U_j)$.

(c) Calculate Evidence

Calculate $P(F_i|U_j)$ as the total probability of observing feature set $F_i$ under usage $U_j$

$$P(F_i|U_j) = \sum_G (P(F_i|G, U_j) \times P(G|U_j))$$

(d) Calculate Posterior Probabilities

Calculate $P(G|F_i, U_j)$ using the Bayes theorem as mentioned above for each usage scenario, considering the relevant feature subset for each: $P(G|F_1, U_1), P(G|F_2, U_2), \cdots, P(G|F_N, U_M)$.

4. Case Study

The objective of this classification task is to predict the quality grade $G \in \{A, B, C, D, E\}$ of prefabricated concrete wall panels based on five input features $F = (F_1, F_2, F_3, F_4, F_5)$ and contextual usage scenario $U$.

According to the sample data listed in Table 1 for a precast concrete wall, the two usge

scenarios are defined as:
$U_1$: External walls of commercial buildings (high-performance requirements: load ⩾ 75%, fire resistance ⩾ 90 min)
$U_2$: Internal walls of warehouses (low-performance requirements: load ⩾ 60%, fire resistance ⩾ 30 min)

Table 1 Sample Data

| Feature | Value | Physical Interpretation |
|---|---|---|
| $F_1$ | 82% | Residual load-bearing capacity |
| $F_2$ | 7mm | Carbonation depth |
| $F_3$ | 0.32W/m²k | Thermal insulation performance |
| $F_4$ | 110min | Fire resistance duration |
| $F_5$ | 12% | Surface damage rate |

The prior probabilities $P(G|U)$ is summairsed in Table 2.

Table 2 Prior Probability for Scenario $U_1$

| Grade | $P(G|U_1)$ | Justification |
|---|---|---|
| A | 0.15 | $U_1$ emphasizes high performance |
| B | 0.25 | |
| C | 0.30 | $U_2$ accepts moderate performance |
| D | 0.20 | |
| E | 0.10 | E-grade rarely accepted in $U_1$ |

To ensure rigorous and consistent probabilistic reasoning, each feature's contribution to the likelihood $P(F \mid G, U)$ is explicitly modeled. Two types of features are considered:
- Continuous Features (e.g., load-bearing capacity, fire resistance): Modeled using Gaussian distributions $N(\mu_{G,U}, \sigma_{G,U})$, where parameters are scenario- and grade-specific.
- Categorical/Bounded Features (e.g., surface damage rate): Modeled using empirically defined probability tables based on proximity to accepted grade thresholds.

To illustrate the approach, we consider scenario $U_1$ and calculate the likelihood for each feature value under each possible grade. summarizes the outcome of these computations, where continuous values have been converted into probabilities using the appropriate normal distribution function.

Table 3 Likelihood Table for Scenario $U_1$

| Feature $F_k$ | G=A | G=B | G=C | G=D | G=E |
|---|---|---|---|---|---|
| $F_1 = 82\%$ | 0.004 ($\mu=90, \sigma=5$) | 0.11 ($\mu=80, \sigma=5$) | 0.07 ($\mu=70, \sigma=10$) | 0.01 ($\mu=60, \sigma=10$) | 0.00 ($\mu=50, \sigma=15$) |
| $F_2 = 7mm$ | 0.8 | 0.6 | 0.4 | 0.2 | 0.1 |
| $F_3 = 0.32$ | 0.7 | 0.9 | 0.5 | 0.3 | 0.1 |
| $F_4 = 110min$ | 0.9 | 0.8 | 0.6 | 0.3 | 0.0 |
| $F_5 = 12\%$ | 0.6 | 0.8 | 0.7 | 0.5 | 0.3 |

Based on the value in Table 3, the following is to compute the joint likelihood for each grade by multiplying the likelihoods of the five features, assuming conditional independence given the grade and scenario.

Grade A:
$$P(F \mid A, U_1) = 0.004 \times 0.8 \times 0.7 \times 0.9 \times 0.6 = 0.0012$$

Grade B:
$$P(F \mid B, U_1) = 0.11 \times 0.6 \times 0.9 \times 0.8 \times 0.8 = 0.0304$$
Grade C:
$$P(F \mid C, U_1) 0.07 \times 0.4 \times 0.5 \times 0.6 \times 0.7 = 0.0059$$

For the likelihood of Grades D and E, the likelihoods are approximately zero, primarily due to violations of critical thresholds in $F_1$ and $F_4$.

The evidence term $P(F \mid U_1)$, serving as the denominator in Bayes' theorem, is obtained by summing the weighted joint likelihoods across all grades:

$$P(F \mid U_1) = 0.0012 \cdot 0.15 + 0.0304 \cdot 0.25 + 0.0059 \cdot 0.30 + \text{(negligible terms)} \approx 0.0089$$

With all necessary components derived, the posterior probabilities $P(G|F, U_1)$ is computed following:

$$P(G|F, U_1) = \frac{P(F|G, U_1) \cdot P(G|U_1)}{P(F|U_1)}$$

Grade A:
$$P(A \mid F, U_1) = \frac{0.0012 \cdot 0.15}{0.0089} = 2\%$$

Grade B:
$$P(B \mid F, U_1) = \frac{0.0304 \cdot 0.25}{0.0089} = 85.4\%$$

Grade C:
$$P(B \mid F, U_1) = \frac{0.0059 \cdot 0.30}{0.0089} = 12.6\%$$

The analysis concludes that, under scenario $U_1$, the sample wall panel is most likely to be classified as Grade B, with a posterior probability of 85.4%. This outcome reflects a high-performance profile across most criteria, albeit marginally below the strict thresholds of Grade A.

Similarly, for scenario ,$U_2$ the likelihood is calculated following the same steps. In contrast to commercial facades, warehouses exhibit more relaxed performance requirements, particularly in terms of structural load and fire resistance. Given the lower performance thresholds of $U_2$ the prior distribution of grades is adjusted accordingly, reflecting the higher acceptability of moderate- to low-performance components in warehouse settings. Table 4 presents the revised prior probabilities. This distribution serves as a reflection of industry tolerance under scenario-specific performance constraints.

Table 4 Prior Probability for Scenario $U_2$

| Grade | $P(G|U_2)$ | Justification |
|---|---|---|
| A | 0.05 | High-performance components rarely required |
| B | 0.20 | Moderate performance generally sufficient |
| C | 0.50 | Most common acceptance grade |
| D | 0.20 | Occasionally accepted with minor repairs |
| E | 0.05 | Poor-performance components rarely used |

Recognizing that lower performance standards are acceptable under $U_2$, the likelihood function $P(F|G, U_2)$ is adjusted accordingly. Notably, features such as load-bearing capacity and fire resistance, which are critical under $U_1$, receive reduced emphasis in $U_2$.

Key adjustments include:

Load-bearing capacity ($F_1$): Lower values are more acceptable under $U_2$, resulting in increased likelihood for grades B and C.

Fire resistance ($F_4$): Given that the minimum threshold for fire resistance is 30 minutes in this scenario, a measured value of 110 minutes significantly exceeds requirements, increasing its likelihood for mid-range grades.

Table 5 Likelihood Table for Scenario $U_2$

| Feature $F_k$ | G=A | G=B | G=C | G=D | G=E |
|---|---|---|---|---|---|
| $F_1 = 82\%$ | 0.004 | 0.9 | 0.8 | 0.4 | 0.1 |
| $F_2 = 7mm$ | 0.6 | 0.9 | 0.6 | 0.3 | 0.1 |
| $F_3 = 0.32$ | 0.5 | 0.7 | 0.6 | 0.4 | 0.2 |
| $F_4 = 110min$ | 0.6 | 0.5 | 0.7 | 0.5 | 0.1 |
| $F_5 = 12\%$ | 0.5 | 0.6 | 0.7 | 0.6 | 0.4 |

These probabilities reflect a relaxed performance evaluation system and contribute to the scenario-specific posterior inference.

The joint likelihood for each grade is computed by multiplying the individual feature likelihoods, assuming conditional independence. For example, for grade C:

$$P(F|C, U_2) = 0.8 \times 0.6 \times 0.6 \times 0.7 \times 0.7 = 0.141$$

Table 6 summarises the results across all grades.

Table 6 Joint likelihoods for all grades under scenario $U_2$

| Grade | $P(F|G, U_2)$ |
|---|---|
| A | 0.7×0.6×0.5×0.6×0.5 = 0.063 |
| B | 0.9×0.7×0.7×0.5×0.6 = 0.132 |
| C | 0.8×0.6×0.6×0.7×0.7 = 0.141 |
| D | 0.4×0.3×0.4×0.5×0.6 = 0.014 |
| E | 0.1×0.1×0.2×0.1×0.4 = 0.000 |

To normalize the posterior probabilities, the evidence term is calculated as the weighted sum of the joint likelihoods, using the scenario-specific priors

$$P(F | U_2) = 0.063 \times 0.05 + 0.132 \times 0.20 + 0.141 \times 0.50 + 0.014 \times 0.20 + 0.000 \times 0.05 = 0.092$$

Finally, the posterior probabilities are calculated by applying Bayes' rule.

Table 7 Posterior probabilities for each grade under scenario $U_2$

| Grade | $P(G|F, U_2)$ |
|---|---|
| A | 0.063×0.05 / 0.092 = 0.034 |
| B | 0.132×0.20 / 0.092 = 0.287 |
| C | 0.141×0.50 / 0.092 = 0.766 |
| D | 0.014×0.20 / 0.092 = 0.030 |
| E | 0.000×0.05 / 0.092 = 0.000 |

The classification result indicates that under the warehouse interior wall scenario $U_2$, the sample panel is most likely to be classified as Grade C, with a posterior probability of 76.6%. This suggests that the panel is directly usable without requiring repair or reinforcement, aligning with the relaxed acceptance criteria for warehouse applications.

To highlight the flexibility and adaptability of the Bayesian classifier, a comparison is drawn

between the outcomes of scenarios $U_1$ and $U_2$ in

Table 8 Scenario comparison summary.

| Aspect | $U_1$: Commercial Exterior Wall | $U_2$: Warehouse Interior Wall |
|---|---|---|
| Most Probable Grade | B (85.4%) | C (76.6%) |
| Repair Requirement | Moderate (minor repair may be needed) | Low (suitable as-is) |
| Dominant Features | Load-bearing, fire resistance | Load-bearing, surface quality |

This comparison in Figure 9 clearly showing how the same panel is more likely to be Grade B under commercial conditions (U₁) and Grade C under warehouse conditions (U₂).

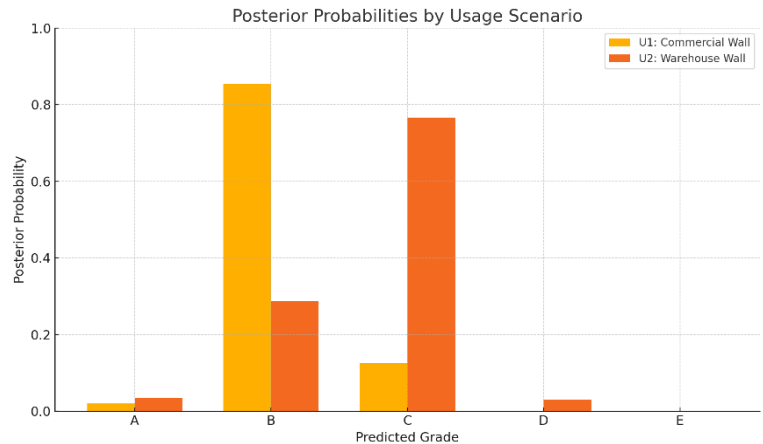

Figure 9 Posterior Probabilities by Usage Scenario

5. Discussion and Conclusion

Th results for the $U_1$ and $U_2$ illustrates the model's capacity to respond to diverse functional contexts, ensuring performance-based classification aligned with practical engineering requirements. The hypothetical confusion matrix in Figure 10, assuming the true grade is C in both cases. It demonstrates that the model might over-predict one level higher (Grade B) under stricter criteria, reflecting a cautious classification stance under scenario $U_1$. This indicates the feature-based model is sensitive to contextual expectations and tends to err on the side of caution when requirements are strict which beneficial in high-risk settings.

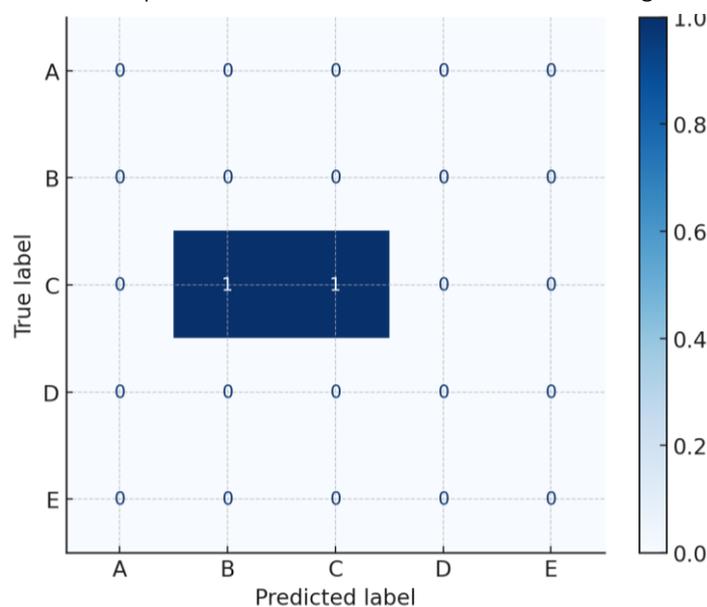

Figure 10 Hypotherical Confusion Matrix for $U_1$ and $U_2$

The Sankey diagram in Figure 11 illustrates the flow of probabilities and scenario influence. This diagram illustrates the flow of classification probability from a synthetic precast concrete panel through two distinct usage scenarios $U_1$ (Commercial Exterior Wall) and $U_2$ (Warehouse Interior Wall). Each scenario leads to a distribution of reuse grades (A–E), reflecting how contextual performance requirements influence probabilistic classification outcomes.

Under $U_1$, stricter demands on load capacity and fire resistance result in a higher likelihood of classification as Grade B, while $U_2$'s more lenient criteria increase the probability of assignment to Grade C. This visualisation enhances explainability by making the relationship between the usage scenario and grade prediction transparent.

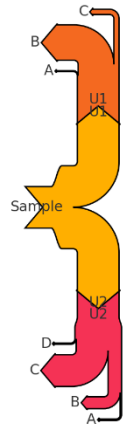

Figure 11 Scenario-Based Grade Classification

The decision tree in Figure 12 maps logical decision paths used for classification. This decision tree represents a rule-based model trained on key features, such as load capacity, carbonation depth, fire resistance, and surface damage, to predict the reuse grade of precast concrete panels.

Each split corresponds to a learned threshold value, supporting engineering interpretability by revealing how different attributes influence classification decisions. The structure allows practitioners to trace the path leading to a specific grade and understand which physical performance measures triggered each decision, thus reinforcing alignment with engineering design logic.

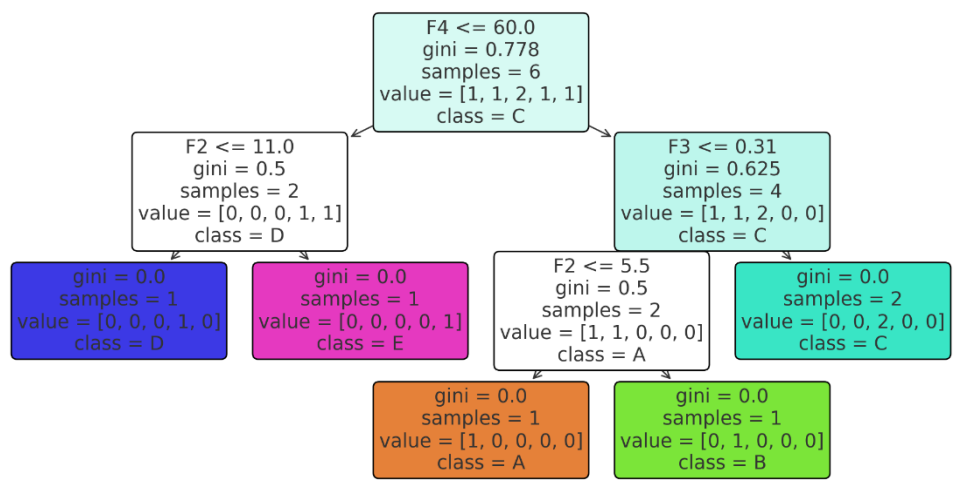

Figure 12 Feature-Based Grade Classification

This study presents an interpretable and adaptive framework for classifying the reuse potential of end-of-life building components. The Multi-Level Grading and Classification System (MGCS) advances the circular economy agenda by integrating Bayesian inference with engineering-specific decision rules to produce scenario-sensitive reuse grades. Its probabilistic reasoning and rule-based logic are designed to predict outcomes and support transparent, auditable decision-making aligned with real-world performance expectations.

Notably, the framework addresses three critical domains of impact: (1) it enhances sustainability by diverting reusable materials from landfill and reducing environmental burden; (2) it improves economic outcomes by preserving material value and supporting cost-effective repair strategies; and (3) it streamlines operational efficacy in EoL handling, enabling more intelligent sorting, planning, and material logistics.

Sankey diagrams depict scenario-driven grade flows, and decision trees clarify feature-based reasoning, strengthening the model's explainability. With future extensions, such as dynamic feature weighting, integration with digital twin systems, and lifecycle impact coupling, the MGCS framework holds promise as a core decision-support tool for scalable, intelligent material reuse in the built environment.

**Acknowledgement:** This research is funded by EPSRC through the Interdisciplinary Circular Economy Centre for Mineral-Based Construction Materials from the UK Research and Innovation (EPSRC Reference: EP/V011820/1)

6. References:

Abbott, GR, JJ McDuling, SA Parsons, and JC Schoeman. 2007. 'Building Condition Assessment: A Performance Evaluation Tool towards Sustainable Asset Management'.

Ajayi, Saheed O., Lukumon O. Oyedele, Muhammad Bilal, Olugbenga O. Akinade, Hafiz A. Alaka, Hakeem A. Owolabi, and Kabir O. Kadiri. 2015. 'Waste Effectiveness of the Construction Industry: Understanding the Impediments and Requisites for Improvements'. *Resources, Conservation and Recycling* 102:101–12.

Akinade, Olugbenga O., Lukumon O. Oyedele, Saheed O. Ajayi, Muhammad Bilal, Hafiz A. Alaka, Hakeem A. Owolabi, Sururah A. Bello, Babatunde E. Jaiyeoba, and Kabir O. Kadiri. 2017. 'Design for Deconstruction (DfD): Critical Success Factors for Diverting End-of-Life Waste from Landfills'. *Waste Management* 60:3–13. doi: 10.1016/j.wasman.2016.08.017.

Anon. 2017. 'ISO 20245:2017 Cross-Border Trade of Second-Hand Goods'.

Anon. n.d.-a. 'Modern Methods of Construction: Guidance for Building Standards Verification'. Retrieved 3 October 2023 (http://www.gov.scot/publications/modern-methods-construction-mmc-guidance-building-standards-verification/).

Anon. n.d.-b. 'Modern Methods of Construction Working Group: Developing a Definition Framework'. *GOV.UK*. Retrieved 3 October 2023 (https://www.gov.uk/government/publications/modern-methods-of-construction-working-group-developing-a-definition-framework).

Anon. n.d.-c. 'Redundant, Repurpose, Rebound'. Retrieved 26 September 2023 (https://www.rics.org/news-insights/redundant-repurpose-rebound--is-it-time-to-reimagine-the-use-of-commercial-real-estate-space).

Anon. n.d.-d. 'Technical_manual_v11'. Retrieved 3 October 2023 (https://www.premierguarantee.com/media/1209/technical_manual_v11_-_chapter_3.pdf).

Bertino, Gaetano, Johannes Kisser, Julia Zeilinger, Guenter Langergraber, Tatjana Fischer, and Doris Österreicher. 2021. 'Fundamentals of Building Deconstruction as a Circular Economy Strategy for the Reuse of Construction Materials'. *Applied Sciences* 11(3):939. doi: 10.3390/app11030939.

Boje, Calin, Annie Guerriero, Sylvain Kubicki, and Yacine Rezgui. 2020. 'Towards a Semantic Construction Digital Twin: Directions for Future Research'. *Automation in Construction* 114:103179. doi: 10.1016/j.autcon.2020.103179.

BSI British Standards. n.d. *Buildings and Constructed Assets. Service-Life Planning*. doi: 10.3403/BSISO15686.

Cai, Gaochuang, and Danièle Waldmann. 2019a. 'A Material and Component Bank to Facilitate

Material Recycling and Component Reuse for a Sustainable Construction: Concept and Preliminary Study'. *Clean Technologies and Environmental Policy* 21(10):2015–32. doi: 10.1007/s10098-019-01758-1.

Cai, Gaochuang, and Danièle Waldmann. 2019b. 'A Material and Component Bank to Facilitate Material Recycling and Component Reuse for a Sustainable Construction: Concept and Preliminary Study'. *Clean Technologies and Environmental Policy* 21(10):2015–32. doi: 10.1007/s10098-019-01758-1.

Cumo, Fabrizio, Federica Giustini, Elisa Pennacchia, and Carlo Romeo. 2022. 'The "D2P" Approach: Digitalisation, Production and Performance in the Standardised Sustainable Deep Renovation of Buildings'. *Energies* 15(18):6689. doi: 10.3390/en15186689.

Faqih, Faisal, and Tarek Zayed. 2021. 'A Comparative Review of Building Component Rating Systems'. *Journal of Building Engineering* 33:101588. doi: 10.1016/j.jobe.2020.101588.

Figl, Hildegund, Carolin Thurner, Frank Dolezal, Patricia Schneider-Marin, and Isabell Nemeth. 2019. 'A New Evaluation Method for the End-of-Life Phase of Buildings'. P. 012024 in *IOP conference series: Earth and environmental science*. Vol. 225. IOP Publishing.

Fivet, Corentin. 2019. 'Design of Load-Bearing Systems for Open-Ended Downstream Reuse'. *IOP Conference Series: Earth and Environmental Science* 225:012031. doi: 10.1088/1755-1315/225/1/012031.

Foster, Gillian, Halliki Kreinin, and Sigrid Stagl. 2020. 'The Future of Circular Environmental Impact Indicators for Cultural Heritage Buildings in Europe'. *Environmental Sciences Europe* 32(1):141. doi: 10.1186/s12302-020-00411-9.

GOV UK. 2022. *Construction Playbook*.

Hořínková, Dita. 2021. 'Advantages and Disadvantages of Modular Construction, Including Environmental Impacts'. P. 032002 in *IOP conference series: Materials science and engineering*. Vol. 1203. IOP Publishing.

Huuhka, S., T. Kaasalainen, J. H. Hakanen, and J. Lahdensivu. 2015. 'Reusing Concrete Panels from Buildings for Building: Potential in Finnish 1970s Mass Housing'. *Resources, Conservation and Recycling* 101:105–21. doi: 10.1016/j.resconrec.2015.05.017.

Iacovidou, Eleni, and Phil Purnell. 2016. 'Mining the Physical Infrastructure: Opportunities, Barriers and Interventions in Promoting Structural Components Reuse'. *Science of The Total Environment* 557–558:791–807. doi: 10.1016/j.scitotenv.2016.03.098.

Khasreen, Mohamad, Phillip F. Banfill, and Gillian Menzies. 2009. 'Life-Cycle Assessment and the Environmental Impact of Buildings: A Review'. *Sustainability* 1(3):674–701. doi: 10.3390/su1030674.

NHBC. n.d. 'Prefabricated-Building-Units'. Retrieved 2 October 2023 (https://www.nhbc.co.uk/binaries/content/assets/nhbc/tech-zone/nhbc-standards/nhbc-accepts/prefabricated-building-units.pdf).

Rakhshan, Kambiz, Jean-Claude Morel, Hafiz Alaka, and Rabia Charef. 2020. 'Components Reuse in the Building Sector – A Systematic Review'. *Waste Management & Research: The Journal for a Sustainable Circular Economy* 38(4):347–70. doi: 10.1177/0734242X20910463.

Salim, NAA, and NF Zahari. 2011. 'Developing Integrated Building Indicator System (IBIS)(a Method of Formulating the Building Condition Rating)'. *Procedia Engineering* 20:256–61.

Sepasgozar, Samad M. E., Felix Kin Peng Hui, Sara Shirowzhan, Mona Foroozanfar, Liming Yang, and Lu Aye. 2020. 'Lean Practices Using Building Information Modeling (BIM) and Digital Twinning for Sustainable Construction'. *Sustainability* 13(1):161. doi: 10.3390/su13010161.

Straub, Ad. 2009. 'Dutch Standard for Condition Assessment of Buildings'. *Structural Survey* 27(1):23–35.

Suchorzewski, Jan, Fabio Santandrea, and Katarina Malaga. 2023. 'Reusing of Concrete Building Elements – Assessment and Quality Assurance for Service-Life'. *Materials Today: Proceedings* S2214785323040658. doi: 10.1016/j.matpr.2023.07.195.

Vanier, DJ "Dana". 2001. 'Why Industry Needs Asset Management Tools'. *Journal of Computing in Civil Engineering* 15(1):35–43.